\documentclass[a4paper,11pt]{article}
\pdfoutput=1 

\usepackage{jheppub} 

\title{\boldmath Long-range correlation of CMB temperature fluctuations from the holographic entanglement entropy}

\author[a]{Hadyan Luthfan Prihadi,}
\author[a]{Donny Dwiputra,}
\author[a,b,1]{Freddy Permana Zen\note{Corresponding author.}}

\affiliation[a]{Theoretical Physics Laboratory, Department of Physics, Institut Teknologi Bandung, Jl. Ganesha 10 Bandung, Indonesia.}
\affiliation[b]{Indonesia Center for Theoretical and Mathematical Physics (ICTMP), Institut Teknologi Bandung, Jl. Ganesha 10 Bandung,
	40132, Indonesia.}
\emailAdd{hadyanluthfanp@s.itb.ac.id}
\emailAdd{donny.dwiputra@s.itb.ac.id}
\emailAdd{fpzen@fi.itb.ac.id}

\abstract{In this paper, we define the holographic multipartite entanglement entropy for $N$ separated subsystems living in a compact $\text{CFT}_d$ space-time. In a large $N$ limit, we find that the first-order holographic entanglement entropy perturbation is proportional to the change in the length of the subsystem and it has symmetric (Gaussian) distribution property. From that finding, we propose to construct two-point correlation functions of holographic entanglement entropy fluctuations analogous to the one that is used in the cosmic microwave background (CMB) temperature fluctuations analysis. Using the first law of thermodynamics, we may correlate tiny changes in entanglement entropy with the temperature fluctuations. By comparing with Planck 2018 CMB data from $j=2$ to $j=2499$ where $j$ denotes the multipole moment, we extract the distribution of the entangling region size that corresponds to the temperature fluctuations. Since the distribution of the entangling region size can be interpreted as the CMB temperature fluctuations, we conclude that entanglement might play a role in the quantum aspects of cosmology.}

\begin{document} 
\maketitle
\flushbottom

\section{Introduction}
\label{sec:1}
In the era of quantum information theory, entanglement is one of the most important phenomena that no classical counterparts could explain, namely, it only appears in the quantum world. Information contained in an entangled state vector $|\psi\rangle$ may be extracted from an entanglement measure, or a measure of how entangled a state $|\psi\rangle$ is. The latter gives rise to the definition of von Neumann entropy as the entanglement entropy that is used in this work to extract the information from an entangled state $|\psi\rangle$.\\
\indent Nowadays, the AdS/CFT correspondence is one of the fastest-growing subjects in theoretical physics \cite{maldacena1997}. It gives a remarkable geometrical method for calculating entanglement entropy using the Ryu-Takayanagi area law or commonly called the holographic entanglement entropy \cite{RyuTakayanagi2006PRL, RyuTakayanagi2006JHEP}. This area law is analogous to the black hole entropy that scales as its horizon area. Furthermore, the extension of holographic entanglement entropy for covariant description has been widely discussed in \cite{Hubeny_2007, Dong_2016, Headrick_2014}.\\
\indent Other holographic entanglement measures such as the holographic entanglement of purification \cite{Umemoto_2018a} and its multipartite counterpart \cite{Umemoto_2018b} have also been discussed. Besides its practicality, there are myriad benefits from the holographic entanglement entropy such as investigating quantum entanglement in inflationary cosmology \cite{Koh_2018}, using entanglement to build space-time \cite{Raamsdonk_2010}, calculating the scrambling time for BTZ black holes \cite{Caputa_2015,Stikonas_2019}, and many more. Other applications of quantum information in cosmological physics can be seen in \cite{Kanno_2015a, Kanno_2015b}. \\
\indent Despite various topics in applying the holographic entanglement entropy to cosmology, analysis of homogeneity, inhomogeneity, and anisotropy of the Universe from the cosmic microwave background (CMB) sky using this prescription is still lacking. However, there are several works that investigate entangled states in CMB sky such as in \cite{Chen_2019}, while the analysis of CMB anisotropy and non-Gaussianity using entropy method are already been discussed in \cite{Minkov_2019, Helling_2006}. The latter shows that Wehrl entropy can be used in CMB analysis. Although the two are different, both Wehrl and von Neumann entropy can be considered as a measure of how a state deviates from a classical prescription, i.e. they measure how "quantum" a state is. This gives us a hint to use the entanglement entropy to analyze CMB anisotropy.\\
\indent Analyzing physical cosmology using quantum mechanics is interesting since, for a long time, scientists always wish to figure out the quantum properties of the early Universe on a large scale. The theory of inflation plays an essential role in investigating the early Universe \cite{Guth_1981}. On the other hand, it is believed that the early Universe shall express quantum mechanical properties as well. For example, quantum fluctuations in primordial Universe has been widely suggested as a perturbation of quantum field in a cosmological background (see \cite{Danielsson_2003, Straumann_2006} and references therein). Furthermore, primordial fluctuations can also be used to analyze the non-Gaussian features of cosmology \cite{Maldacena_2003}. Another conceptual problem in cosmology is how the primordial quantum fluctuations in the early Universe undergo quantum-to-classical transition such that it appears classical to us \cite{kiefer2009cosmological}.\\
\indent In this work, we propose how investigations of holographic entanglement entropy fluctuations can be used to analyze homogeneity, inhomogeneity, and anisotropy of the Universe, especially for CMB sky observation, using two-point correlation functions. We suggest that the information about homogeneity, inhomogeneity and anisotropy of the Universe may be extracted from initially-entangled quantum states via its entanglement measure, i.e. its von Neumann entropy. Moreover, this information-storing entangled states might be the quantum memory pattern suggested by Gia Dvali \cite{Dvali_2019, Dvali_2018a, Dvali_2018b}. In relating entropy with temperature, the theory of thermodynamics should play an important role in this work.\\
\indent Quantum field theory is of particular interest in the study of entanglement entropy in many-body systems, both in ground and low-energy excited states. As the quantum field possesses an infinite number of degrees of freedom, its vacuum is highly entangled. The entanglement temperature rising from the holographic entanglement is related to the usual thermal energy in low-energy gapless CFT systems \cite{swingle2013universal,chen2015universal}. There are vast research literature studying the thermodynamics aspects of entanglement from entropy perturbation, such as \cite{Nadi_2019, Bhattacharya_2013, Saha_2019, Wong_2013, Alishahiha_2013, Laskhari_2014} (for the extended entanglement thermodynamics, see \cite{Kastor_2016}). In the AdS/CFT correspondence language, perturbing the entropy means adding the pure AdS geometry with small metric perturbation. Since the entanglement entropy shares similar properties with the thermodynamical counterpart, such as the positivity of the relative entropy, many suggest \cite{Nadi_2019, Bhattacharya_2013, Saha_2019, Wong_2013, Alishahiha_2013, Kastor_2016, Laskhari_2014} that entanglement obeys similar laws with the ones in thermodynamics. The first law is analogous to the $TdS$ equation and it is widely discussed that the change of entanglement entropy may correspond to the change of modular Hamiltonian \cite{Casini_2011, Laskhari_2014}. In this work, we use the analog of $TdS$ equation to match the change of entanglement entropy with temperature fluctuation and use is to analyze the CMB sky. \\ 
\indent The structure of this work is as follows. We begin the work in Sec. \ref{sec:2} by studying a highly-entangled quantum memory state that is scrambled by a phenomena called the holographic jumps. The definition of entanglement entropy for a particular isolated subsystem with respect to the rest and its extension to define the multipartite entanglement entropy is discussed in Sec. \ref{sec:3} along with some holographic calculations for $\text{AdS}_3/\text{CFT}_2$ circular disk and $\text{AdS}_{d+1}/\text{CFT}_d$ straight belt systems. In Sec. \ref{sec:4}, the use of the holographic multipartite entanglement entropy for spherical symmetric two-point correlation functions for the entropy fluctuations is discussed while the extension for CMB anisotropy analysis is investigated in Sec. \ref{sec:5} along with the interpretation to the observed Planck 2018 CMB data. To sum up, we provide some conclusions and further discussions for this work in Sec. \ref{sec:6}.
\section{Entangled Patterns of Quantum Memories}
\label{sec:2}
We begin this work by studying a highly entangled quantum states and its origin. This section is a review based on the idea of quantum memory pattern suggested by Gia Dvali \cite{Dvali_2019,Dvali_2018a,Dvali_2018b}. It is shown in \cite{Dvali_2019}, that the Universe might have a primordial memory pattern which is stored in a de Sitter quantum state. A large number of information is stored in a Fock space,
\begin{equation}
|\psi\rangle=|pattern\rangle = |\psi_1 \psi_2, ..., \psi_N\rangle,\label{eq:pattern}
\end{equation}  
that is defined as the memory space, where $N\in\mathbb{N}$ is some large number.\\
\indent This pattern is an eigenstate of a Hamiltonian that is consisted of number operators $\hat{a}_k^{\dagger}\hat{a}_k$,
\begin{equation}
\hat{H}=\sum_{k}\mathcal{E}_k\hat{a}_k^{\dagger}\hat{a}_k,
\end{equation}
where $\{\mathcal{E}_k\}$ are the energy modes, $\{\hat{a}_k,\hat{a}_k^{\dagger}\}$ are the annihilation and creation operators, and $k=1,2,...,N$. As mentioned in \cite{Dvali_2019}, there is a critical point, $N_0$, that when $\langle \hat{a}_0^{\dagger}\hat{a}_0\rangle=N_0$, the energy gap $\mathcal{E}_k$ becomes zero; this is called the gapless mode. Following this assumption, the Hamiltonian is now can be written in the form of
\begin{equation}
\hat{H}=\sum_{k\neq 0}\varepsilon_k\bigg(1-\frac{\hat{a}_0^{\dagger}\hat{a}_0}{N_0}\bigg)\hat{a}_k^{\dagger}\hat{a}_k+\varepsilon_0\hat{a}_0^{\dagger}\hat{a}_0.
\end{equation}
\indent An existence of the first term indicates that the Hamiltonian has an enhanced memory capacity \cite{Dvali_2018b, Dvali_2019} (see also \cite{Dvali_2019b}) since in the critical point, the energy gap becomes zero and an enormous amount of information can be excited into a quantum memory pattern, such as in Eq. (\ref{eq:pattern}), without providing any energy as a cost. This is called the assisted gaplessness since the zero mode $\hat{a}_0^{\dagger}\hat{a}_0$, as the master mode, determines the gaplessness of the Hamiltonian \cite{Dvali_2019}. One may recognize that if $\langle\hat{a}_0^{\dagger}\hat{a}_0\rangle$ deviates from $N_0$, the energy gap also moves away from gaplessness and hence it is more difficult to store information in the excited states. This phenomenon is known as the memory burden effect and in some literatures (such as in \cite{Dvali_2018b}), this explains how a quantum memory pattern is prevented to decay and gets stabilized.\\
\indent There are possibilities that some other critical number, $N_0'\neq N_0$, which present a gaplessness in the Hamiltonian exist. Moreover, even though the state is kept by its own memory burden, there is a non-zero probability for the state to decay into some other modes, say $\hat{b}_k$, that is commutes with all $\hat{a}_k$ modes, until $\langle\hat{a}^{\dagger}_0\hat{a}_0\rangle$ reaches another critical point $N_0'$. It is possible because it can be assumed that this happens right after the 'graceful exit' described in \cite{Dvali_2019} and the process produces an entangled memory state. This particular event is then investigated as the holographic jumps \cite{Dvali_2018b} that deserve further discussions.
\subsection{The Holographic Jumps}
\label{sec:2.1}
Suppose that now the modes $\hat{a}_k$ may decay into other external modes $\hat{b}_k$ that are commute with all $\hat{a}_k$ modes. The external modes $\hat{b}_k$ does not necessarily to be an enhanced memory capacity. The Hamiltonian is then provides an extra interaction terms between $\hat{a}_k$ and $\hat{b}_k$, and hence can be written as
\begin{align}
\label{eq:Hamiltonian1}\hat{H}=&\sum_{k\neq 0}\varepsilon_k\bigg(1-\frac{\hat{a}_0^{\dagger}\hat{a}_0}{N_0}\bigg)\hat{a}_k^{\dagger}\hat{a}_k+\varepsilon_0\hat{a}_0^{\dagger}\hat{a}_0\\\nonumber
&\;\;\;\;\;+\sum_k\varepsilon_k\hat{b}_k^{\dagger}\hat{b}_k+\frac{1}{2N_0}\varepsilon_0(\hat{a}_0^{\dagger}\hat{b}_0+\hat{b}_0^{\dagger}\hat{a}_0)+...\;,
\end{align}
where we ignore all higher-mode interaction terms since we only need the mode $\hat{b}_0$ as a reservoir. The coupling constant is normalized to $\frac{1}{2N_0}$ for convenience following \cite{Dvali_2019}. One may see that the factor $\frac{1}{2N_0}$ determines the decay probability amplitude. If $N_0$ is very large, then the decay will be less likely to happen (but not zero) since the crossing term between $\hat{a}_0$ and $\hat{b}_0$ is heavily suppressed. In primordial de Sitter era point of view \cite{Dvali_2019}, both $N$ and $N_0$ represent the same quantity, i.e. the Gibbons-Hawking entropy \cite{Gibbons_1977}, and it is taken to be a large number$^1$\note{This also follows from \cite{Dvali_2018a} in the double-scaling limit prescription. On the other hand, $N\rightarrow\infty$ is also known as the semi-classical limit \cite{Dvali_2017, Dvali_2016}.} that determines the quantum break-time of de Sitter \cite{Dvali_2017}. This limit is also important in our work and will be discussed more in Sec. \ref{sec:3}.\\
\indent The Hamiltonian in the form of Eq. (\ref{eq:Hamiltonian1}) does not have any gapless mode other than $N_0$. If we want to introduce another gapless mode, for example $N_0'=N_0-\Delta N_0$, the Hamiltonian is then written as
\begin{align}
\label{eq:Hamiltonian2}\hat{H}&=\sum_{k\in k_1}\varepsilon_k\bigg(1-\frac{\hat{a}_0^{\dagger}\hat{a}_0}{N_0}\bigg)\hat{a}_k^{\dagger}\hat{a}_k+\sum_{k\in k_2}\varepsilon_k\bigg(1-\frac{\hat{a}_0^{\dagger}\hat{a}_0}{N_0-\Delta N_0}\bigg)\hat{a}_k^{\dagger}\hat{a}_k+\varepsilon_0\hat{a}_0^{\dagger}\hat{a}_0\\\nonumber
&\;\;\;\;\;+\sum_k\varepsilon_k\hat{b}_k^{\dagger}\hat{b}_k+\frac{1}{2N_0}\varepsilon_0(\hat{a}_0^{\dagger}\hat{b}_0+\hat{b}_0^{\dagger}\hat{a}_0)+...\;.
\end{align}
This Hamiltonian is defined such then $k_1$ be the first gapless modes where $k_2$ be the second. If at initial stage all the pattern is stored in $k_1$ mode while the zero mode fully consists of $N_0$, 
\begin{equation}
|in\rangle=|N_0\rangle_{a_0}\otimes|0\rangle_{b_0}\otimes|pattern\rangle_{k_1}\otimes|0\rangle_{k_2},
\end{equation}
after the decay process reaches $\Delta N_0$ in $\hat{b}_0$ mode, we will have a final state that can be written as
\begin{equation}
|out\rangle=|N_0-\Delta N_0\rangle_{a_0}\otimes|\Delta N_0\rangle_{b_0}\otimes|0\rangle_{k_1}\otimes|pattern'\rangle_{k_2}.
\end{equation}
\indent It is important to notice that $|in\rangle\rightarrow|out\rangle$ process rises possibilities for the pattern state in $k_1$ be completely rewritten into a new quantum memory of $k_2$ since zero energy is needed in order to excite the information into $k_2$ modes after the decay reaches $\Delta N_0$ in $\hat{b}_0$ state. Now, all of the information is stored in the new memory-storing pattern, $|pattern'\rangle_{k_2}$. This process is called the holographic jumps \cite{Dvali_2018b}. The term "holographic" is used in the literature since it is shown in \cite{Dvali_2018a,Dvali_2018b} that the microstate entropy of such pattern follows the area-law. However, the latter mentioned entropy is not an entanglement entropy and hence we are not discussing them further.\\
\indent The difference between $|pattern\rangle_{k_1}$ and $|pattern'\rangle_{k_2}$ will be discussed hereafter. From investigations in \cite{Dvali_2018b}, the rewriting process scrambles the memory space and hence the state $|pattern'\rangle_{k_2}$ becomes highly entangled, or it can be written as
\begin{equation}
|pattern'\rangle_{k_2}=\sum_{\psi_1...\psi_N}C_{\psi_1,...,\psi_N}|\psi_1,...,\psi_N\rangle,
\end{equation}
where $C_{\psi_1,...,\psi_N}$ is some coefficients that determine how entangled the state $|pattern'\rangle_{k_2}$ is. This procedure will be important for the main purpose of this work since we investigate a highly-entangled quantum memory state. In general, there might be more than two gapless modes. For instance, the jumps might occur several times until it finally reaches certain point, say $k_e$, with $e>2$. The exact number of $e$ is not really important since what we need is an entangled memory state, and the entanglement measure is later computed by the von Neumann entropy.\\
\indent There are several aspects of this process that are worth mentioning such as the time needed to rewrite all of the memory pattern into the new one and other Hamiltonian possibilities. The former gives rise to further investigation of a time evolution functions of $|in\rangle\rightarrow|out\rangle$ process. These functions can be obtained by time-evolving $|in\rangle$ state with Hamiltonian in Eq. (\ref{eq:Hamiltonian2}). The other form of Hamiltonian that might be possible is
\begin{equation}
\hat{H}=\sum_{k\neq 0}\varepsilon_k\bigg(1-\frac{\hat{a}_0^{\dagger}\hat{a}_0}{N_0}\bigg)^m\hat{a}_k^{\dagger}\hat{a}_k+\varepsilon_0\hat{a}_0^{\dagger}\hat{a}_0+...\;,
\end{equation}
for some positive integer $m$ that corresponds to the memory burden \cite{Dvali_2019}. However, those aspects do not yet need further investigations in this work.
\section{Holographic Multipartite Entanglement Entropy}
\label{sec:3}
Before we dig deeper into the main aspect of this paper, we would like to review the computation of entanglement entropy along with its holographic duals and expand them to define the multipartite entanglement entropy for the purposes of this work. Moreover, behavior of the entanglement entropy due to small area perturbation for $\text{AdS}_3/\text{CFT}_2$ circular system and $\text{AdS}_{d+1}/\text{CFT}_d$ straight belt system are also investigated.

\subsection{Definition of Multipartite Entanglement Entropy}
\label{sec:3.1}
Let $L=\bigcup_{i=1}^N A_i^{(j)}$ be a (d-1)-dimensional spatial region in a constant time-slice of a $\text{CFT}_d$ that consists of $N\in\mathbb{N}$ number of causally disconnected regions $A_1^{(j)},...,A_N^{(j)}$ for some integer $j$ and each $A_i^{(j)}$, $i=1,...,N$, belongs to some Hilbert space $\mathcal{H}_{A_i^{(j)}}$ with a quantum state vector $|\psi_i\rangle^{(j)}$. The need for an index $j$ will be clear after discussing the analysis of a spherically symmetric two-point correlation function.\\
\indent Now let $|\psi\rangle^{(j)}$ be an entangled state that is defined by
\begin{equation}
|\psi\rangle^{(j)}=\sum_{\psi_1...\psi_N} C_{\psi_1,...,\psi_N}|\psi_1\rangle^{(j)}\otimes...\otimes|\psi_N\rangle^{(j)}\equiv\sum_{\psi_1...\psi_N}C_{\psi_1,...,\psi_N}|\psi_1,...,\psi_N\rangle^{(j)},
\end{equation}
where $C_{\psi_1,...,\psi_N}$ is some coefficients which play a role in determining how entangled a state $|\psi\rangle^{(j)}$\note{ Notice that $|\psi\rangle^{(j)}$ is the highly-entangled quantum memory pattern that is defined in Sec. \ref{sec:1}.} is$^2$. We may assume that $|\psi\rangle^{(j)}$ is in a ground state and $N$ is a large number. From that, we are able to construct a pure density matrix $\rho_{A_1^{(j)}...A_N^{(j)}}\equiv|\psi\rangle^{(j)(j)}\langle\psi|=:\rho^{(j)}$ and a reduced one $\rho_{A_i^{(j)}}\equiv\text{Tr}_{A_i^{(j)c}}\rho^{(j)}$ where $A_i^{(j)c}$ is the complement of $A_i^{(j)}$. The region $A_i^{(j)}$ is said to be entangled with respect to the rest if $\rho_{A_i^{(j)}}$ is mixed. In this work, we use the entanglement entropy, i.e. von Neumann entropy, defined by
\begin{equation}
S_{A_i^{(j)}}=-\text{Tr}\rho_{A_i^{(j)}}\log\rho_{A_i^{(j)}},
\end{equation}
for the entanglement measure, or the measure of how mixed a density matrix $\rho_{A_i^{(j)}}$ is. \\
\indent The multipartite entanglement entropy for a state $|\psi\rangle^{(j)}$ is defined as
\begin{equation}
M^{(j)}=\sum_{i=1}^{N}S_{A_i^{(j)}}.\label{eq:3}
\end{equation}
This definition comes from the multipartite entanglement of purification in \cite{Umemoto_2018b},
\begin{equation}
\Delta_P(A_1^{(j)}:...:A_N^{(j)})=\sum_{i=1}^NS_{A_i^{(j)}}-S_{A_1^{(j)}...A_N^{(j)}},
\end{equation}
with $\rho_{A_1^{(j)}...A_N^{(j)}}$ is taken to be pure, i.e. $S_{A_1^{(j)}...A_N^{(j)}}=0$ and therefore no purification is needed. Even though it is called the multipartite entanglement entropy, it still measures the sum of usual entanglement entropy for a bipartite system since it traces out all other spaces except one, for instance, $A_i^{(j)}$. Nevertheless, this definition is valuable for advance investigation done in this paper. The definition of multipartite entanglement entropy in Eq. (\ref{eq:3}) also comes from \cite{Alishahiha_2015}, related to the mutual information.\\ 
\indent Explicit calculation of the entanglement entropy for a certain system is ordinarily complicated in quantum field theory language \cite{Calabrese_2004, Calabrese_2009}. Fortunately, there is a holographic method proposed by Ryu and Takayanagi \cite{RyuTakayanagi2006PRL,RyuTakayanagi2006JHEP} that calculates the entanglement entropy geometrically using the $\text{AdS/CFT}$ correspondence.
\subsection{Holographic Multipartite Entanglement Entropy}
\label{sec:3.2}
\begin{figure}
	\begin{center}
		\includegraphics[scale=0.7]{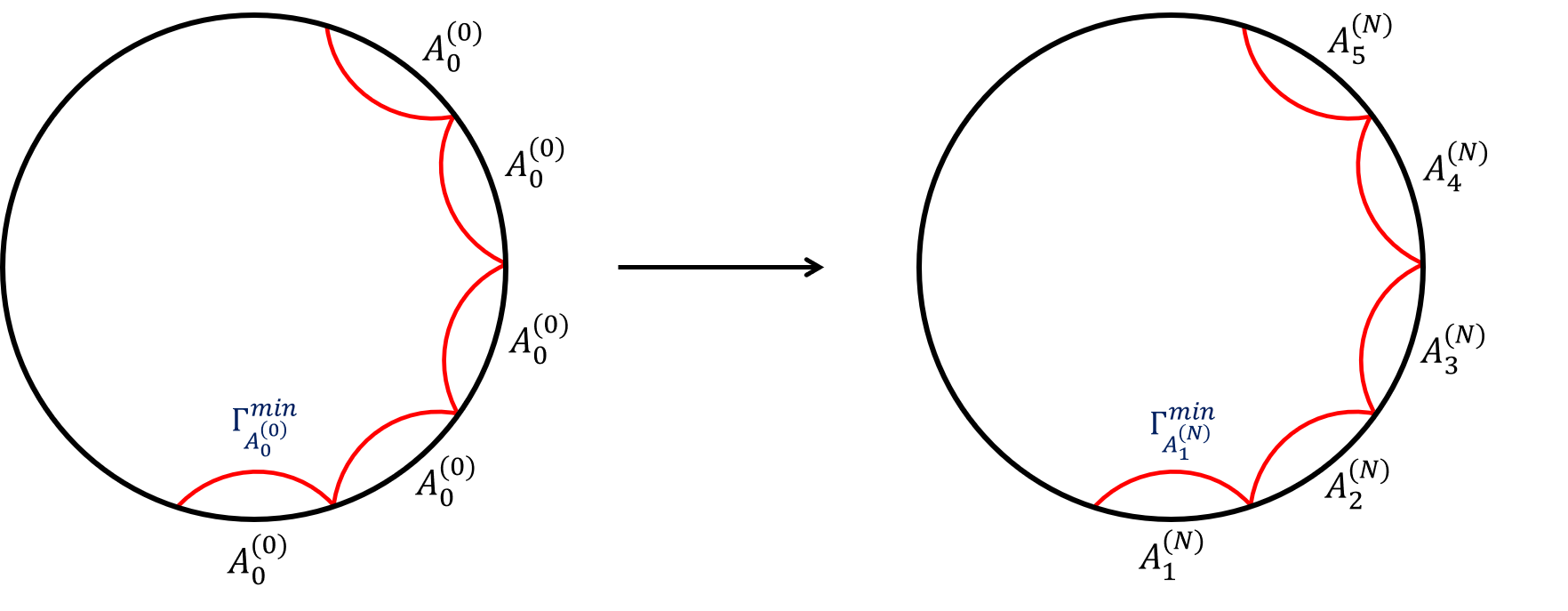}\caption{Those images present the model realization in $\text{AdS}_3/\text{CFT}_2$ circular system in a constant time slice. $\text{AdS}_3$ is described by the whole disk while the $\text{CFT}_2$ is its boundary depicted by the circumference. Minimized regions are the red curved lines, defined as $\Gamma_{A_i^{(j)}}$. A system in the left figure is the homogeneous mode with $j=0$ while the right one describes multipole mode with $j=N$. Both systems have equal circumference, $L$, even though the subsystems' area change. What important is, both system still have equal number of subsystems, which is $N$.}\label{figure:1}
	\end{center}
\end{figure}

\indent In holographic point of view, using Ryu-Takayanagi area law \cite{RyuTakayanagi2006PRL,RyuTakayanagi2006JHEP}, the entanglement entropy can be computed by calculating its minimum entangling surface in the bulk. Consider a d-dimensional quantum field theory ($\text{CFT}_d$) in the boundary of a (d+1)-dimensional Anti de Sitter space-time ($\text{AdS}_{d+1}$) and all regions $A_1^{(j)},...,A_N^{(j)}$ live in a constant time slice of a $\text{CFT}_d$ space-time. Suppose that the region $A_i^{(j)}$ is separated with the rest by a horizon as the boundary $\partial A_i^{(j)}$. Hence, there exist some region in the bulk associated with $A_i^{(j)}$, denoted by $\Gamma_{A_i^{(j)}}$, that satisfies
\begin{equation}
\partial\Gamma_{A_i^{(j)}}=\partial A_i^{(j)}.
\end{equation}
\indent A region that minimizes the area of $\Gamma_{A_i^{(j)}}$ is defined as $\Gamma_{A_i^{(j)}}^{\text{min}}$ and hence its corresponding spatial (d-1)-dimensional area in $\text{AdS}_{d+1}$ can be written as $\text{Area}(\Gamma_{A_i^{(j)}}^{\text{min}})$. As a result, the holographic entanglement entropy can be written as
\begin{equation}
S_{A_i^{(j)}}=\frac{\text{Area}(\Gamma_{A_i^{(j)}}^{\text{min}})}{4G_N^{(d+1)}},
\end{equation}
where $G_N^{(d+1)}$ is the Newton's gravitational constant in (d+1) dimension. Following this relation, we can write the holographic dual of the multipartite entanglement entropy, or HMEE for short, as
\begin{equation}
M^{(j)}=\sum_{i=1}^{N}\frac{\text{Area}(\Gamma_{A_i^{(j)}}^{\text{min}})}{4G_N^{(d+1)}}.
\end{equation}
\indent For further purposes in this work, we would like to redefine the multipartite entanglement entropy as follows. In the monopole case with $j=0$, $M^{(0)}$ shall represents a homogeneous mode with a uniform entanglement entropy distribution and hence we propose that in this case, all areas are equal, i.e. $\text{Area}(A_1^{(0)})=\text{Area}(A_2^{(0)})=...=\text{Area}(A_N^{(0)})\equiv \text{Area}(A^{(0)}_0)$$^3$\note{the term area in $\text{Area}(A_i^{(j)})$ can be misleading since in $\text{CFT}_d$ point of view, $\text{Area}(A_{i}^{(j)})$ is actually a spatial volume of a region $A_{i}^{(j)}$. However, we want to make clear that $\text{Area}(A_{i}^{(j)})$ is the (d-1)-dimensional area in the boundary of $\text{AdS}_{d+1}$ space-time.}. Henceforth, the multipartite entanglement entropy can be written as
\begin{equation}
M^{(0)}=NS_{A_0^{(0)}}\equiv N\frac{\text{Area}(\Gamma_{A_0^{(0)}}^{\text{min}})}{4G_N^{(d+1)}}.
\end{equation}
\indent In the dipole case with $j=1$, there should be three, i.e. $2j+1$ with $j=1$, different areas; $\text{Area}(A_{-1}^{(1)}),\text{Area}(A_0^{(1)}),\text{Area}(A_1^{(1)})$ and there are $N/3$ regions with each area. Hence, the multipartite entanglement entropy can be written as
\begin{equation}
M^{(1)}=\frac{N}{3}S_{A_{-1}^{(1)}}+\frac{N}{3}S_{A_0^{(1)}}+\frac{N}{3}S_{A_1^{(1)}}.
\end{equation}
\indent Furthermore, we can extend to the quadrupole and even higher $j$ cases until we have $2j+1$ different areas. Following those relations, we can generically define the multipartite entanglement entropy as
\begin{equation}
M^{(j)}=\frac{N}{2j+1}\sum_{m=-j}^{j}S_{A_m^{(j)}}=\frac{N}{2j+1}\sum_{m=-j}^{j}\frac{\text{Area}(\Gamma_{A_m^{(j)}}^{\text{min}})}{4G_N^{(d+1)}}.
\end{equation}
As a result, in the limit $\lim_{j\rightarrow0}M^{(j)}=NS_{A_{0}^{(0)}}$, it reduces to the homogeneous case while in the limit $\lim_{j\rightarrow N}M^{(j)}\simeq\sum_{m=1}^{N}S_{A_m^{(N)}}$, it becomes the original definition of multipartite entanglement entropy. In other words, $N$ can be interpreted as the maximum number for $j$ as a cutoff, or $N=2j_{\text{max}}+1$ for some large integer $j_{\text{max}}$.\\
\indent Since $\rho_{A_1^{(j)}...A_N^{(j)}}$ is needed to be pure, a boundary condition
\begin{equation}
\frac{N}{2j+1}\sum_{m=-j}^{j}\text{Area}(A_m^{(j)})=\text{Area}(L)=N\text{Area}(A_0^{(0)}),\label{eq:12}
\end{equation}
needs to be satisfied. It can be seen that $L\equiv\bigcup_{m=-j_{\text{max}}}^{j_{\text{max}}}A_m^{(j)}$ and $S_{L}=S_{A_1^{(j)}...A_N^{(j)}}=0$ for all $j$. $\text{Area}(L)$ can be interpreted as the total area of a compact $\text{CFT}_d$. This condition can easily be understood using a geometric prescription, such as in Fig. \ref{figure:1}, and somehow can be interpreted as normalization condition. In general, $L$ can be different for each $j$, i.e. $L\equiv L^{(j)}$. However in this work, we demand that $L$ is identical for all $j$.$^4$
\note{To avoid confusion, throughout this paper, $j$ is used to define the angular moment index with $j\simeq\frac{\pi}{\theta}$ like in the usual CMB analysis. On the other hand, $l$ is used to define the size of the entangling surface area; in this case, it is denoted as $l_i^{(j)}$.}
\subsection{Compact Circular System in $\text{AdS}_3/\text{CFT}_2$}
\label{sec:3.3}
\begin{figure}
	\begin{center}
		\includegraphics[scale=1]{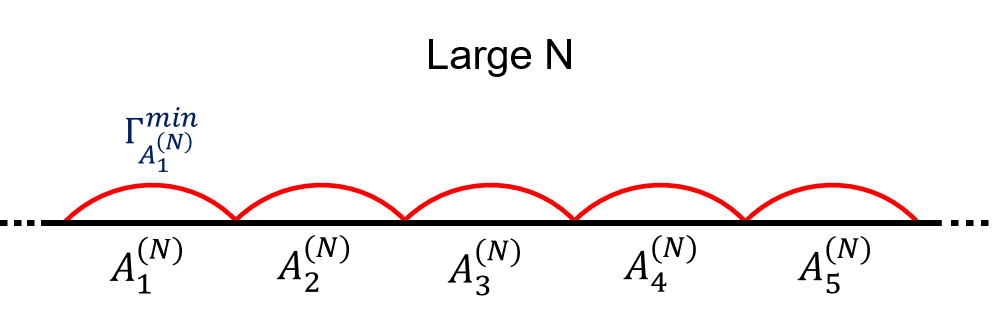}\caption{This image shows how a circular system is flattened out into a 1-dimensional straight one for large $N$. This approximation is useful since we can extend it into a higher dimensional system. In this limit, we still have $N$ number of subsystems, where $N$ is very large.}\label{figure:2}
	\end{center}
\end{figure}

In this section, some explicit calculations for the HMEE in a 2-dimensional circular system of $\text{CFT}_2$ with circumference $L$ are shown (The correct notation for the circumference should be $\text{Area}(L)$ but in this section we will write $L$ to make it short). Let $A_i^{(j)}:=\{x|x\in[a_i^{(j)},b_i^{(j)}]\}$, $i=1,...,N$ be one of the subsystems for some integer $j$ in a constant time-slice of a $\text{CFT}_2$ which lives in the boundary of $\text{AdS}_{3}$. We have $b_i^{(j)}=a_{i+1}^{(j)}$, $b_N^{(j)}=a_1^{(j)}$ and in polar coordinate, $0\leq\theta\leq \frac{2\pi}{L}$. \\
\indent Entanglement entropy for subsystem $A_i^{(j)}$ with respect to the rest is given by \cite{RyuTakayanagi2006PRL,RyuTakayanagi2006JHEP}
\begin{equation}
S_{A^{(j)}_i}=\frac{c}{3}\log\bigg(\frac{L}{\pi\epsilon}\sin\bigg(\frac{\pi l^{(j)}_i}{L}\bigg)\bigg),
\end{equation}
where $c$ is a central charge for $\text{CFT}_2$, $\epsilon\rightarrow0$ is a UV cutoff and $l_i^{(j)}\equiv b_{i}^{(j)}-a_i^{(j)}=:\text{Area}(A_i^{(j)})$. The central charge can be interchanged with an $\text{AdS}_3$ radius $R$ via $c=\frac{3R}{2G_N^{(3)}}$ \cite{RyuTakayanagi2006JHEP,RyuTakayanagi2006PRL} to switch between $\text{CFT}_2$ calculation and its holographic dual. Furthermore, the boundary condition in Eq. (\ref{eq:12}) implies that
\begin{equation}
\frac{N}{2j+1}\sum_{m=-j}^{j}l_m^{(j)}=L\;\;\;\;\;\rightarrow\;\;\;\;\;\frac{1}{2j+1}\sum_{m=-j}^{j}\delta l_m^{(j)}=0,
\end{equation}
where $\delta l_m^{(j)}$ is the change of every $\text{Area}(A_m^{(j)})$.\\
\indent In monopole or homogeneous mode with $j=0$, we only have one area,
\begin{equation}
\text{Area}(A_0^{(0)})=l_0^{(0)}=L/N,
\end{equation}
and therefore the HMEE is given by
\begin{equation}
M^{(0)}=\frac{Nc}{3}\log\bigg(\frac{L}{\pi\epsilon}\sin\bigg(\frac{\pi l^{(0)}_0}{L}\bigg)\bigg)=\frac{Nc}{3}\log\bigg(\frac{L}{\pi\epsilon}\sin\bigg(\frac{\pi}{N}\bigg)\bigg).
\end{equation}
If $N$ is assumed to be very large, then this reduces to
\begin{equation}
M^{(0)}\simeq\frac{Nc}{3}\log\bigg(\frac{l_0^{(0)}}{\epsilon}\bigg)=\frac{Nc}{3}\log\bigg(\frac{L}{N\epsilon}\bigg).\label{eq:20}
\end{equation}
This is equal to the entanglement entropy for an infinite 1-dimensional system, such as in Fig. \ref{figure:2}. This approximation is useful in this paper since, in this limit, we can easily extend it into a d-dimensional system in $\text{CFT}_d$ space-time which will be discussed further in the next chapter.\\
\indent For the dipole case with $j=1$, we have $l_{-1}^{(1)},l_0^{(1)},l_1^{(1)}$ which satisfy
\begin{equation}
l_{-1}^{(1)}+l_0^{(1)}+l_1^{(1)}=\frac{3L}{N}.
\end{equation}
For example, suppose that
\begin{align}
l_0^{(1)}&=\frac{L}{N},\label{eq:18}\\
l_{-1}^{(1)}&=\frac{L-\Delta L}{N},\label{eq:19}\\
l_1^{(1)}&=\frac{L+\Delta L}{N},\label{eq:21}
\end{align}
where $\Delta L<L$ is the change in area with respect to the homogeneous mode and it is not necessarily to be infinitesimal. Therefore, the HMEE is given by
\begin{align}
M^{(1)}&=\frac{Nc}{9}\biggl\{\log\bigg(\frac{L}{\pi\epsilon}\sin\bigg(\frac{\pi}{N}-\frac{\Delta L}{NL}\bigg)\bigg)+\log\bigg(\frac{L}{\pi\epsilon}\sin\bigg(\frac{\pi}{N}\bigg)\bigg)\\\nonumber
&\;\;\;\;\;\;\;\;\;\;+\log\bigg(\frac{L}{\pi\epsilon}\sin\bigg(\frac{\pi}{N}+\frac{\Delta L}{NL}\bigg)\bigg)\biggl\}\nonumber\\
&=\frac{Nc}{9}\log\bigg[\bigg(\frac{L}{\pi\epsilon}\bigg)^3\sin\bigg(\frac{\pi}{N}-\frac{\Delta L}{NL}\bigg)\sin\bigg(\frac{\pi}{N}\bigg)\sin\bigg(\frac{\pi}{N}+\frac{\Delta L}{NL}\bigg)\bigg].
\end{align}
\indent However, if $\Delta L/N$ is assumed to be very small and we ignore $\mathcal{O}((\Delta L/N)^2)$ and higher, we will have linear form that can be written as
\begin{equation}
M^{(1)}_{\text{lin.}}\simeq\frac{Nc}{3}\log\bigg(\frac{L}{\pi\epsilon}\sin\bigg(\frac{\pi}{N}\bigg)\bigg)=:M^{(0)}.
\end{equation}
This shows that $M^{(1)}$ is somehow immune to small perturbations and it reduces to the homogeneous case $M^{(0)}$ due to the boundary condition in Eq. (\ref{eq:12}). \\
\indent If we take the quadratic term into account, the HMEE will have an extra term and can be written as
\begin{equation}
M^{(1)}_{\text{quad.}}\simeq\frac{Nc}{9}\log\bigg[\bigg(\frac{L}{\pi\epsilon}\bigg)^3\sin\bigg(\frac{\pi}{N}\bigg)\bigg(\sin^2\bigg(\frac{\pi}{N}\bigg)-\bigg(\frac{\Delta L}{NL}\bigg)^2\bigg)\bigg].
\end{equation}
This expression is somehow difficult to be written as $M^{(1)}_{\text{quad.}}=M^{(0)}+\delta M^{(1)}$ since it is still trapped inside a logarithmic function.\\
\indent Finally, we analyze how tiny perturbations affect the multipartite entanglement entropy in large mode where $j\rightarrow j_{\text{max}}$, i.e. $N\rightarrow 2j_{\text{max}}+1$, for some large number $j_{\text{max}}$. In this limit, all areas $\text{Area}(A_m^{(N)})$ are different. Assume that areas from $A_1^{(N)}$ to $A_{N-1}^{(N)}$ changes by $\delta l^{(N)}_1$ to $\delta l^{(N)}_{N-1}$ respectively and to fulfill the boundary condition, the area of $A_N^{(N)}$ needs to change by $\sum_{i=1}^{N-1}-\delta l_i^{(N)}$. Following previous formulation, it is important to make sure that $\delta l_i^{(j)}\propto\mathcal{O}(1/N)$. The HMEE then becomes
\begin{align}
M^{(N)}&=\sum_{m=1}^{N-1}\frac{c}{3}\log\bigg(\frac{L}{\pi\epsilon}\sin\bigg(\frac{\pi(L/N+\delta l_m^{(N)})}{L}\bigg)\bigg)\\\nonumber
&\;\;\;\;\;\;\;\;\;\;+\frac{c}{3}\log\bigg(\frac{L}{\pi\epsilon}\sin\bigg(\frac{\pi(L/N-\sum_{m=1}^{N-1}\delta l_m^{(N)})}{L}\bigg)\bigg)\\\nonumber
&=\frac{c}{3}\log\bigg(\bigg(\frac{L}{\pi\epsilon}\bigg)^{N}\sin\bigg(\frac{\pi(L/N-\sum_{m=1}^{N-1}\delta l_m^{(N)})}{L}\bigg)\prod_{m=1}^{N-1}\sin\bigg(\frac{\pi(L/N+\delta l_m^{(N)})}{L}\bigg)\bigg).
\end{align}
\indent Unfortunately, this expression also reduces to the homogeneous mode with circumference $L$ if all of the quadratic and higher terms of the area perturbations are ignored. It indicates that for $j\rightarrow j_{\text{max}}$, the HMEE is also immune to infinitesimal perturbations. From those findings, we may conclude that for small perturbations, the HMEE is only determined by the $\text{CFT}_2$ total size, $L$, and reduces to the homogeneous mode for all $j$.
\subsection{Large N Straight Belt Approximation in $\text{AdS}_{d+1}/\text{CFT}_d$}
\label{sec:3.4}
In this section, we wish to compute the HMEE in a d-dimensional straight belt system of a $\text{CFT}_d$. As can be seen from Eq. (\ref{eq:20}), for a large system with $N\rightarrow\infty$, i.e. $L\rightarrow\infty$ to keep $l_0^{(0)}$ as a relatively small constant, we can approximate the compact $\text{CFT}_d$ system with an infinite d-dimensional $\text{CFT}_d$ with straight belt subsystems $A_{1}^{(j)}...A_N^{(j)}$. In Poincar\'e coordinate, the $\text{AdS}_{d+1}$ metric associated with this system is given by
\begin{equation}
ds^2_{\text{AdS}_{d+1}} =\frac{R^2}{z^2}\bigg(-dt^2+dz^2+\sum_{i=1}^{d-1}(dx_i)^2\bigg),
\end{equation}
where $R$ is the $\text{AdS}_{d+1}$ radius. The $\text{CFT}_{d}$ lives in the $\text{AdS}_{d+1}$ boundary, denoted as $\lim_{z\rightarrow0}ds^2_{\text{AdS}_{d+1}}=ds^2_{\text{CFT}_d}$.\\
\indent Now let $A^{(j)}_1,...,A^{(j)}_N$ be disjoint regions in this geometry with $A^{(j)}_i:=\{x|x_1\in[a_i^{(j)},b_i^{(j)}],\\x_{2,3,...,d-1}\in[-L/2,L/2]\}$, where we still have $l_i^{(j)}\equiv b_i^{(j)}-a_i^{(j)}$. The holographic entanglement entropy for a subsystem $A_i^{(j)}$ with respect to the rest in this system is given by \cite{RyuTakayanagi2006JHEP,RyuTakayanagi2006PRL}
\begin{equation}
S_{A_i^{(j)}}=\frac{1}{4G_N^{(d+1)}}\bigg[\frac{2R^{d-1}}{d-2}\bigg(\frac{L}{\epsilon}\bigg)^{d-2}-\frac{2^{d-1}\pi^{\frac{(d-1)}{2}}R^{d-1}}{d-2}\bigg(\frac{\Gamma\big(\frac{d}{2(d-1)}\big)}{\Gamma\big(\frac{1}{2(d-1)}\big)}\bigg)^{d-1}\bigg(\frac{L}{l_i^{(j)}}\bigg)^{d-2}\bigg],
\end{equation}
and hence the multipartite entanglement entropy is given by
\begin{equation}
M^{(j)}=\frac{N}{4G_N^{(d+1)}}\bigg[\frac{2R^{d-1}}{d-2}\bigg(\frac{L}{\epsilon}\bigg)^{d-2}-\frac{2^{d-1}\pi^{\frac{d-1}{2}}R^{d-1}}{(d-2)(2j+1)}\bigg(\frac{\Gamma\big(\frac{d}{2(d-1)}\big)}{\Gamma\big(\frac{1}{2(d-1)}\big)}\bigg)^{d-1}\sum_{m=-j}^{j}\bigg(\frac{L}{l_m^{(j)}}\bigg)^{d-2}\bigg].\label{eq:32}
\end{equation}
\indent Without any change in $\text{Area}(A_0^{(0)})$, the multipartite entanglement entropy for monopole or homogeneous mode can be obtained from substituting $l_0^{(0)}=L/N$ into Eq. (\ref{eq:32}) and hence it can be written as
\begin{equation}
M^{(0)}=\frac{N}{4G_N^{(d+1)}}\bigg[\frac{2R^{d-1}}{d-2}\bigg(\frac{L}{\epsilon}\bigg)^{d-2}-\frac{2^{d-1}\pi^{\frac{d-1}{2}}R^{d-1}}{d-2}\bigg(\frac{\Gamma\big(\frac{d}{2(d-1)}\big)}{\Gamma\big(\frac{1}{2(d-1)}\big)}\bigg)^{d-1}(N)^{d-2}\bigg].\label{eq:29}
\end{equation}
This expression gives $M^{(0)}=NS_{A_0^{(0)}}$ and indeed expressing the homogeneous mode.\\
\indent For the dipole mode, using previous assumption for the area changes, we will have
\begin{align}
M^{(1)}=\frac{N}{4G_N^{(d+1)}}\bigg[&\frac{2R^{d-1}}{d-2}\bigg(\frac{L}{\epsilon}\bigg)^{d-2}-\frac{2^{d-1}\pi^{\frac{d-1}{2}}R^{d-1}}{3(d-2)}\bigg(\frac{\Gamma\big(\frac{d}{2(d-1)}\big)}{\Gamma\big(\frac{1}{2(d-1)}\big)}\bigg)^{d-1}\bigg(\frac{NL}{L-\Delta L}\bigg)^{d-2}\\\nonumber
&-\frac{2^{d-1}\pi^{\frac{d-1}{2}}R^{d-1}}{3(d-2)}\bigg(\frac{\Gamma\big(\frac{d}{2(d-1)}\big)}{\Gamma\big(\frac{1}{2(d-1)}\big)}\bigg)^{d-1}(N)^{d-2}\\\nonumber
&-\frac{2^{d-1}\pi^{\frac{d-1}{2}}R^{d-1}}{3(d-2)}\bigg(\frac{\Gamma\big(\frac{d}{2(d-1)}\big)}{\Gamma\big(\frac{1}{2(d-1)}\big)}\bigg)^{d-1}\bigg(\frac{NL}{L+\Delta L}\bigg)^{d-2}\bigg].
\end{align}
From a straightforward Taylor expansion calculation and if we ignore the order of $\mathcal{O}(\Delta L^2)$ and higher, we can see that
\begin{equation}
M^{(1)}_{\text{lin.}}\simeq\frac{N}{4G_N^{(d+1)}}\bigg[\frac{2R^{d-1}}{d-2}\bigg(\frac{L}{\epsilon}\bigg)^{d-2}-\frac{2^{d-1}\pi^{\frac{d-1}{2}}R^{d-1}}{d-2}\bigg(\frac{\Gamma\big(\frac{d}{2(d-1)}\big)}{\Gamma\big(\frac{1}{2(d-1)}\big)}\bigg)^{d-1}(N)^{d-2}\bigg]=:M^{(0)}.
\end{equation}
Again, all of the $\Delta L$ terms are canceling with each other and it reduces to the homogeneous mode. \\
\indent However, if we take the quadratic term into account, we will have
\begin{align}
M^{(1)}_{\text{quad.}}\simeq\frac{N}{4G_N^{(d+1)}}\bigg[&\frac{2R^{d-1}}{d-2}\bigg(\frac{L}{\epsilon}\bigg)^{d-2}-\frac{2^{d-1}\pi^{\frac{d-1}{2}}R^{d-1}}{d-2}\bigg(\frac{\Gamma\big(\frac{d}{2(d-1)}\big)}{\Gamma\big(\frac{1}{2(d-1)}\big)}\bigg)^{d-1}\\\nonumber
&\times(N)^{d-2}\bigg(1-\frac{(d-1)(d-2)}{3}\bigg(\frac{\Delta L}{L}\bigg)^2\bigg)\bigg].
\end{align}
Now we can write the HMEE as a second order perturbation $M_{\text{quad.}}^{(1)}=M^{(0)}+\delta M_{\text{quad.}}^{(1)}$ with
\begin{equation}
\delta M^{(1)}_{\text{quad.}}\equiv\frac{N}{4G_N^{(d+1)}}\frac{(d-1)2^{d-1}\pi^{\frac{d-1}{2}}R^{d-1}}{3}\bigg(\frac{\Gamma\big(\frac{d}{2(d-1)}\big)}{\Gamma\big(\frac{1}{2(d-1)}\big)}\bigg)^{d-1}(N)^{d-2}\bigg(\frac{\Delta L}{L}\bigg)^2,
\end{equation}
where the first-order term vanishes. This gives important insights that when $j$ increases, the HMEE also changes just like it is applied with some area perturbations that affect the entanglement entropies. Even though it only appears after the quadratic term is included, analysis of the linear perturbation is useful for this work.\\
\indent Lastly, we investigate the HMEE for $j\rightarrow j_{\text{max}}$ in this system. Following the same argument in $\text{AdS}_3/\text{CFT}_2$ example for the area perturbations, we will have the HMEE in a form of
\begin{align}
M^{(N)}=\frac{N}{4G_N^{(d+1)}}\bigg[&\frac{2R^{d-1}}{d-2}\bigg(\frac{L}{\epsilon}\bigg)^{d-2}-\frac{2^{d-1}\pi^{\frac{d-1}{2}}R^{d-1}}{N(d-2)}\bigg(\frac{\Gamma\big(\frac{d}{2(d-1)}\big)}{\Gamma\big(\frac{1}{2(d-1)}\big)}\bigg)^{d-1}\sum_{m=1}^{N-1}\bigg(\frac{L}{L/N+\delta l_m^{(j)}}\bigg)^{d-2}\\\nonumber
&-\frac{2^{d-1}\pi^{\frac{d-1}{2}}R^{d-1}}{N(d-2)}\bigg(\frac{\Gamma\big(\frac{d}{2(d-1)}\big)}{\Gamma\big(\frac{1}{2(d-1)}\big)}\bigg)^{d-1}\bigg(\frac{L}{L/N-\sum_{m=1}^{N-1}\delta l_m^{(j)}}\bigg)^{d-2}\bigg],
\end{align}
and also reduces to $M^{(0)}$ if we ignore quadratic terms and higher. \\
\indent A generalization to $\text{AdS}_{d+1}/\text{CFT}_d$ calculation is important since in this limit, we may write the entanglement entropy as a first order perturbation $S_{A_m^{(j)}}=S_{A_0^{(0)}}+\delta S_{A_m^{(j)}}$ easily with
\begin{equation}
\delta S_{A_m^{(j)}}=\frac{2^{d-1}\pi^{\frac{d-1}{2}}R^{d-1}}{4G_N^{(d+1)}}L^{d-2}\bigg(\frac{\Gamma\big(\frac{d}{2(d-1)}\big)}{\Gamma\big(\frac{1}{2(d-1)}\big)}\bigg)^{d-1}\delta l_m^{(j)},
\end{equation} 
and we have the relation $\delta S_{A_m^{(j)}}\propto\delta l_m^{(j)}$. Our boundary condition in Eq. (\ref{eq:12}) implies that the distribution of $\delta l_m^{(j)}$, i.e. the distribution of $\delta S_{A_m^{(j)}}$, over $m$ has to be symmetric and a Gaussian distribution is a perfect match for this. As a consequence, we will always have $\langle \delta l_m^{(j)}\rangle =\langle \delta S_{A_m^{(j)}}\rangle=0$ where $\langle . \rangle$ denotes the averages over $m$,
\begin{equation}
\langle \delta l_m^{(j)}\rangle\equiv\frac{1}{2j+1}\sum_{m=-j}^j\delta l_m^{(j)}.
\end{equation}
\indent One may assume that even though this expression is calculated for a straight belt system, it
will also apply for general systems satisfying a boundary condition in Eq. (\ref{eq:12}) despite further investigations are needed. 
\section{HMEE for Spherical Symmetric Two-Point Correlation Functions}
\label{sec:4}
Imagine a 2-dimensional spherical surface in a particular constant time-slice. Suppose that there is an area element $dA$ in some direction denoted by a unit vector $\hat{n}$ which currently does not have any causal contact with the outside region of $dA$. However, the information contained in a quantum state $|\psi(\hat{n})\rangle$ inside the region $dA$ might be initially entangled with the rest. Therefore, we are able to associate an entanglement entropy to it, denoted with $S(\hat{n})$, as the entanglement measure of $|\psi(\hat{n})\rangle$.\\
\indent Suppose that the entanglement entropy is distributed along a 2-dimensional spherical surface of the Universe with average value over all direction is given by $S\equiv\langle S(\hat{n})\rangle=S_{A_0^{(0)}}$. If there are tiny fluctuations, i.e. the distribution of entanglement entropy of the Universe is not completely homogeneous and isotropic throughout the surface, denoted as $\delta S(\hat{n})\equiv(S(\hat{n})-S)$, it can be expressed as the sum of spherical harmonics functions
\begin{equation}
\delta S(\hat{n})=\sum_{j=0}^{j_{\text{max}}}\sum_{m=-j}^{j}a_{jm}^{S}Y_{jm}(\hat{n}),\label{eq:37}
\end{equation}
where $\{a_{jm}^{S}\}$ are the expansion coefficients for entanglement entropy fluctuations and $\{Y_{jm}(\hat{n})\}$ are the spherical harmonics functions. Hence, the average value of this function should be zero, i.e. $\langle\delta S(\hat{n})\rangle=0$. Following this formulation, the HMEE will be used to analyze this entanglement entropy fluctuation using first order perturbation.\\
\indent In some literature, $\delta S(\hat{n})$ is called the renormalized entanglement entropy, such as in \cite{Saha_2019}, since it does not include any divergent term with $\mathcal{O}(1/\epsilon)$. The reason why $\delta S(\hat{n})$ is used instead of $\frac{\delta S}{S}(\hat{n})$ is also arise from the first law of thermodynamics that is investigated further in Sec. 5.
\subsection{First Order Entropy Perturbation}
\label{sec:4.1}
Due to the previous assumption saying that there are tiny fluctuations of the entanglement entropy distribution, it is virtuous to study the first order entropy perturbation. In the linear order, the HMEE will have the form of
\begin{equation}
M^{(j)}=NS_{A_0^{(0)}}+\frac{N}{2j+1}\sum_{m=-j}^{j}\delta S_{A_m^{(j)}}\equiv NS_{A_0^{(0)}}+N\langle \delta S_{A_m^{(j)}} \rangle.\label{eq:38}
\end{equation}
In other words, the change of the multipartite entanglement entropy is given by $\delta M^{(j)}=N\langle\delta S_{A_m^{(j)}}\rangle\propto N\langle\delta l_m^{(j)}\rangle$. Namely, our definition of the multipartite entanglement entropy is just the statistical average of the entanglement entropy over $m$ with some normalization constant $N$.\\
\indent However, due to the boundary condition imposed in Eq. (\ref{eq:12}), the average value of entropy fluctuations will always be zero. Another approach is to take the root-mean-square value, $\langle\delta S_{A_m^{(j)}}\rangle \equiv \langle \delta S_{A_m^{(j)}}\rangle_{\text{RMS}}$. If we assume that the distribution of $\delta S_{A_m^{(j)}}$ is Gaussian, then the root-mean-square will be the standard deviation of the distribution, $\sigma^{(j)}\equiv C_j^{1/2}$. It is clear that from our definition of the multipartite entanglement entropy, we are able to analyze first order perturbation of the entanglement entropy fluctuation statistically and use it to explore more about two-point correlation functions which will be discussed next.\\
\indent Since the spherical harmonics satisfy orthonormality condition $\int Y_{jm}Y_{j'm'}^*d^2n=\delta_{jj'}\delta_{mm'}$, the inverse relation for Eq. (\ref{eq:37}) can be written as
\begin{equation}
a_{jm}^{S}=\int\delta S(\hat{n})Y_{jm}^*d^2n.\label{eq:40}
\end{equation}
From the similarity of its statistical properties such as the distribution of $a_{jm}^{S}$ should be symmetric and hence the average over $m$ is zero, i.e. $\langle a_{jm}^{S}\rangle=0$, and because from Eq. (\ref{eq:40}) we have $a_{jm}^{S}\propto\delta S(\hat{n})$, we may write
\begin{equation}
a_{jm}^{S}=\delta S_{A_m^{(j)}}. \label{eq:41}
\end{equation}
\indent Nevertheless, the functions $\{a_{jm}\}$ are generally complex while the entanglement entropy perturbation $\delta S_{A_m^{(j)}}$ is always real and positive. Since $\delta S(\hat{n})$ is a real function as well, and $Y_{jm}^*=(-)^mY_{j,-m}$, we have $a_{jm}^{S*}=(-)^ma_{j,-m}^{S}$ and hence,
\begin{equation}
\delta S_{A_m^{(j)}}=(-)^m\delta S_{A_{-m}^{(j)}}.\label{eq:42}
\end{equation} 
This condition, however, does not violates the total area boundary condition in Eq. (\ref{eq:12}), as can be seen from the dipole mode example in Eqs. (\ref{eq:18}-\ref{eq:21}) that exactly matches Eq. (\ref{eq:42}).
\subsection{Statistical Interpretation for Two-Point Correlation Functions}
\label{sec:4.2}
Suppose that we want to know how a fluctuation of entanglement entropy $\delta S(\hat{n})$ in some direction $\hat{n}$ correlates with the other one in a different direction $\hat{m}$. The best way to do that is to write the two-point correlation function of $\delta S(\hat{n})$ as
\begin{align}
\langle\delta S(\hat{n})\delta S(\hat{m})\rangle&=\sum_{j=0}^{j_{\text{max}}}\frac{(2j+1)}{4\pi}\langle |a_{jm}^{S}|^2\rangle P_j(\hat{n}\cdot\hat{m})\\\nonumber
&=\sum_{j=0}^{j_{\text{max}}}\frac{(2j+1)}{4\pi}\langle \delta S_{A_m^{(j)}}\delta S_{A_m^{(j)}}\rangle P_j(\hat{n}\cdot\hat{m}),
\end{align}
where $\{P_j(\hat{n}\cdot\hat{m})\}$ are the Legendre polynomials. In this point of view, one can obtain the distribution variance as
\begin{equation}
\langle\delta S_{A_m^{(j)}}\delta S_{A_m^{(j)}}\rangle = \sigma^{(j)2}=C_j.\label{eq:4.7}
\end{equation}
\indent An interesting result from this formulation is that we can analyze the distribution of entanglement entropy fluctuations throughout the Universe, via the two-point correlation functions, using the previously defined multipartite entanglement entropy. Also, in this formulation, a divergent term with $\mathcal{O}(1/\epsilon)$ that is contained in $S_{A_0^{(0)}}$ is not present.\\
\indent The Universe is assumed to be statistically homogeneous and hence, the two-point correlation functions of the entanglement entropy perturbation, now defined as 
\begin{equation}
\Delta(\hat{n},\hat{m})\equiv\langle\delta S(\hat{n})\delta S(\hat{m})\rangle,
\end{equation}
will be a function of the separation distance $r$ between $\hat{n}$ and $\hat{m}$, where $r\equiv|\hat{n}-\hat{m}|$, only. This means $\Delta(\hat{n},\hat{m})=\Delta(r)$ and the power spectrum $\mathcal{P}(k)$ may be extracted using Fourier transform of $\Delta(r)$,
\begin{equation}
\mathcal{P}(k)=2\pi\int_{0}^{\infty}\frac{\sin(kr)}{kr}\Delta(r)r^2 dr,
\end{equation}
where the power spectrum $\mathcal{P}(k)$ is a function of a wave number $k$. This gives another alternative to analyze the two-point correlation function for entanglement entropy perturbation analogous to the analysis of matter density inhomogeneities in the Universe.\\ 
\indent If the distribution of entanglement entropy fluctuation $\langle\delta S_{A_m^{(j)}}\rangle$ related to the one in the CMB temperature anisotropy, then we can also conclude that the CMB temperature distribution, either its homogeneity (when all $j$(s) vanishes except for homogeneous mode $j=0$) or inhomogeneity and anisotropy (when all other $j\neq0$ are included) is affected by quantum entanglement. Moreover, information about non-Gaussianity might also be extracted from the distribution, nevertheless, that is beyond the scope of this study.\\
\indent Before continuing to more detailed investigations about CMB sky, general expressions for entanglement entropy via area perturbation due to the change of a metric in the bulk space-time will be discussed.
\subsection{General Expressions for Area Perturbation}
\label{sec:4.3}
Previously, the change in area of a subsystem $A$, $\delta \text{Area}(A)$, is investigated without bothering what kind of metric perturbation in the bulk causes it. Now we would like to present a general expression of an area perturbation $\delta \text{Area}(A)$ that gives rise to a change in the holographic entanglement entropy
\begin{equation}
\delta S_{A}=\frac{\delta\text{Area}(\Gamma^{\text{min}}_{A})}{4 G_N^{(d+1)}},
\end{equation}
as a result of a metric perturbation in $\text{AdS}_{d+1}$ bulk space-time.\\
\indent In this section, we will follow similar calculation for area variation done in \cite{Nadi_2019}. Suppose that there is a (d+1)-dimensional anti de Sitter space-time $\text{AdS}_{d+1}$ with an unperturbed metric $g_{MN}(x)$, where $M,N=0,1,...,d$. Since we want to measure the entanglement entropy, then we need to investigate how $\text{Area}(\Gamma_{A}^{\text{min}})$ change if the metric $g_{MN}$ is perturbed. Suppose that $h_{ab}(\zeta)$, with $a,b=0,1,...,d-1$, is an induced metric in the bulk that is defined as
\begin{equation}
h_{ab}=g_{MN}\partial_a x^{M}\partial_b x^{N},
\end{equation}
where $\partial_a\equiv\frac{\partial}{\partial\zeta^a}$ and $h\equiv\det(h_{ab})$. Hence, the area of $\Gamma_A$ can be expressed as
\begin{equation}
\text{Area}(\Gamma_A)=\int d^{d-1}\zeta\sqrt{h}.
\end{equation}
\indent To determine its minimum area, $\text{Area}(\Gamma_A)$ needs to be varied. However, it is shown in \cite{Nadi_2019} that in order to get the geodesic, $h_{ab}$ needs to satisfy a so-called Gauss equation,
\begin{equation}
\partial_a(\sqrt{h}h^{ab}\partial_bx^{M})+\sqrt{h}h^{ab}\Gamma^{M}_{NK}\partial_ax^{N}\partial_bx^{K}=0,\label{eq:45}
\end{equation}
where $\Gamma^{M}_{NK}$ is the Christoffel symbol for the metric $g_{MN}$. Equipped with an induced metric $h_{ab}$ that satisfies Eq. (\ref{eq:45}), an arbitrary $\text{Area}(\Gamma_{A})$ becomes the geodesic, $\text{Area}(\Gamma_A^{\text{min}})$.\\
\indent At last, we arrive in formulating $\delta\text{Area}(\Gamma_A^{\text{min}})$, or the change of minimum area due to external metric perturbation in the bulk. Suppose that now the total variation of the metric $g_{MN}$ come from both trajectory and external variation,
\begin{equation}
\delta g_{MN}=\frac{\partial g_{MN}}{\partial x^{K}}\delta x^{K}+g_{MN}^{(1)}.
\end{equation}
The minimum area variation is then given by
\begin{equation}
\delta\text{Area}(\Gamma_A^{\text{min}})=\frac{1}{2}\int d^{d-1}\zeta\sqrt{h}h^{ab}(g_{MN}^{(1)}\partial_a x^M \partial_b x^N),
\end{equation}
and hence,
\begin{equation}
\delta S_A=\frac{1}{8G_N^{(d+1)}}\int d^{d-1}\zeta\sqrt{h}h^{ab}(g_{MN}^{(1)}\partial_a x^M \partial_b x^N).
\end{equation}
\indent Usually, the metric perturbation comes from the existence of a black hole in the bulk and it directly affects $\Gamma_A$. This prescription also means that the system is now excited. However, in this work, we view that the change in the area of $A$ affects the change in the entanglement entropy via $\delta\text{Area}(\Gamma_A^{\text{min}})$ in response to a metric perturbation in the bulk. Although it is not known yet what causes the perturbation of $g_{MN}$, in general, it does not have to be a black hole in the bulk; it can be any arbitrary gravitational perturbation.
\section{Two-Point Correlation Functions for CMB Anisotropy Analysis}
\label{sec:5}
The analysis of two-point correlation functions for CMB temperature distribution is one of the most important studies in cosmology. In this section, we examine how the two-point correlation functions of the entanglement entropy perturbation can be used to analyze the CMB temperature fluctuations $\frac{\delta T}{T}(\hat{n})\equiv\frac{T(\hat{n})-T}{T}$, where $T\equiv\langle T(\hat{n})\rangle$ is the average over all space. To relate temperature and entropy fluctuations, $TdS$ equation is used as the first law of entanglement thermodynamics.
\subsection{TdS Equation From the First Law of Entanglement Thermodynamics}
\indent It has been shown (see, for example, \cite{Nadi_2019,Bhattacharya_2013,Saha_2019,Wong_2013,Alishahiha_2013,Kastor_2016}) that entanglement entropy provides some relations that is analogous to thermodynamics, or it is called "entanglement thermodynamics" in \cite{Alishahiha_2013, Laskhari_2014} (see also \cite{swingle2013universal, chen2015universal}). Following those prescriptions, relation between the change of entropy $\delta S$ and temperature fluctuation $\delta T$ may rise from the first law of thermodynamics via $TdS$ equation,
\begin{equation}
\delta S = C_V \frac{\delta T}{T}+\mathcal{K}T\frac{\delta V}{T},
\end{equation}
where $C_V$ is the heat capacity at constant volume, and $\mathcal{K}$ is some constant that depends on the system. The second term is a change of volume in $\text{CFT}_d$, i.e. the area variation in $\text{AdS}_{d+1}$. It is proportional to $L^{d-2}\delta l$ and hence also satisfy $\delta V \propto \delta S$. We may absorb the proportionality constant into $\mathcal{K}$ and write the equation in the form of
\begin{equation}
\delta S = \mathcal{C_V}\frac{\delta T}{T},
\end{equation}
where $\mathcal{C_V}$ is a new constant that depends on $C_V$, $\mathcal{K}$, and dimension. The temperature $T$ should be called the entanglement temperature and $\mathcal{C_V}$ should be defined as the entanglement heat capacity.
\subsection{Two-Point Correlation Function of Temperature Fluctuation}
\indent The relation between $\delta S(\hat{n})$ and $\frac{\delta T}{T}(\hat{n})$ has been investigated and now we proceed to construct a two-point correlation function for $\frac{\delta T}{T}(\hat{n})$. The two-point correlation function for two different directions $\hat{n}$ and $\hat{m}$ is then
\begin{align}
\bigg\langle\frac{\delta T}{T}(\hat{n})\frac{\delta T}{T}(\hat{m})\bigg\rangle&\propto\langle\delta S(\hat{n})\delta S(\hat{m})\rangle\\\nonumber
&\propto\sum_{j=0}^{j_{\text{max}}}\frac{(2j+1)}{4\pi}\langle \delta S_{A_m^{(j)}}\delta S_{A_m^{(j)}}\rangle P_j(\hat{n}\cdot\hat{m}),
\end{align}
where the proportionality constant is just $\mathcal{C_V}^{-2}$ in the right hand side. We have to relate $\delta S$ with $\frac{\delta T}{T}$ since the available data for CMB analysis is in the form of temperature fluctuations. From the previous analysis, we are able to compute the temperature two-point function from the distribution variance $C_j$ of entanglement entropy fluctuations.\\
\indent For example, in the dipole mode with $j=1$, using Eq. (\ref{eq:4.7}) we have
\begin{equation}
C_1=\frac{2}{3}\delta S_{A_1^{(1)}}^2.
\end{equation}
The variance $C_1$ only depends on one entropy perturbation, $\delta S_{A_1^{(1)}}$. In a d-dimensional straight-belt system language, the dipole variance can be written as
\begin{equation}
C_1=\frac{2^{2d-3}}{3G_N^{(d+1)}}\pi^{d-1}R^{2(d-1)}L^{2d-4}\bigg(\frac{\Gamma\big(\frac{d}{2(d-1)}\big)}{\Gamma\big(\frac{1}{2(d-1)}\big)}\bigg)^{2(d-1)} (\delta l_1^{(1)})^2.
\end{equation}
If we take $\delta l_1^{(1)}=\Delta L/N$ and $l_0^{(0)}=L/N$, we have
\begin{equation}
C_1=\frac{2^{2d-3}}{3G_N^{(d+1)}}\pi^{d-1}R^{2(d-1)}L^{2d-6}(l_0^{(0)})^2\bigg(\frac{\Gamma\big(\frac{d}{2(d-1)}\big)}{\Gamma\big(\frac{1}{2(d-1)}\big)}\bigg)^{2(d-1)} (\Delta L)^2.
\end{equation}
\indent Since $L$ is taken to approach infinity, if $d<3$, $C_1$ is suppressed by the factor of $\frac{1}{L^{6-2d}}$ and hence it will infinitesimally small and conversely, $C_1$ diverges if $d>3$ since it will be enhanced by the factor of $L^{2d-6}$. Therefore, the good choice for the dimension is $d=3$ and besides, it is the most suitable number for our use since we investigate distributions on a 2-dimensional surface. After taking $d=3$, we have
\begin{equation}
C_1=\frac{8\pi^2 R^4}{3G_N^{(3)}}\bigg(\frac{\Gamma\big(\frac{3}{4}\big)}{\Gamma\big(\frac{1}{4}\big)}\bigg)^{4} (l_0^{(0)})^2(\Delta L)^2.\label{eq:62}
\end{equation}
\indent From Eq. (\ref{eq:62}), it can be concluded that the dipole mode $C_1$ in $d=3$ depends on two parameters: $l_0^{(0)}$ and $\Delta L$. However, we may correlate $l_0^{(0)}$ with the average temperature of CMB and hence the degree of freedom is reduced by one. Another point to be remarked is that the choice of the dimension $d$ is quite arbitrary since it only depends on how the state $|\psi\rangle^{(j)}$ stores its information.
\subsection{Distribution of entangling surface that matches the CMB data}
Up to this point, we have shown that geometric entanglement between subsystems in the early universe may be responsible for the correlation of two space-like separated points. We have found that the first-order hologrophic entanglement entropy ($d=3$) is proportional to the length $L$ of the subsystems. Thus, the anisotropy of CMB power spectrum, which represents the seeds of galaxies, is directly related to variation of $L$. We match the distribution of $\Delta L$ to satisfy the actual CMB last scattering surface using the Legendre transform in polar coordinates.

\begin{figure}[h]	
	\centering
	\includegraphics[width=\linewidth]{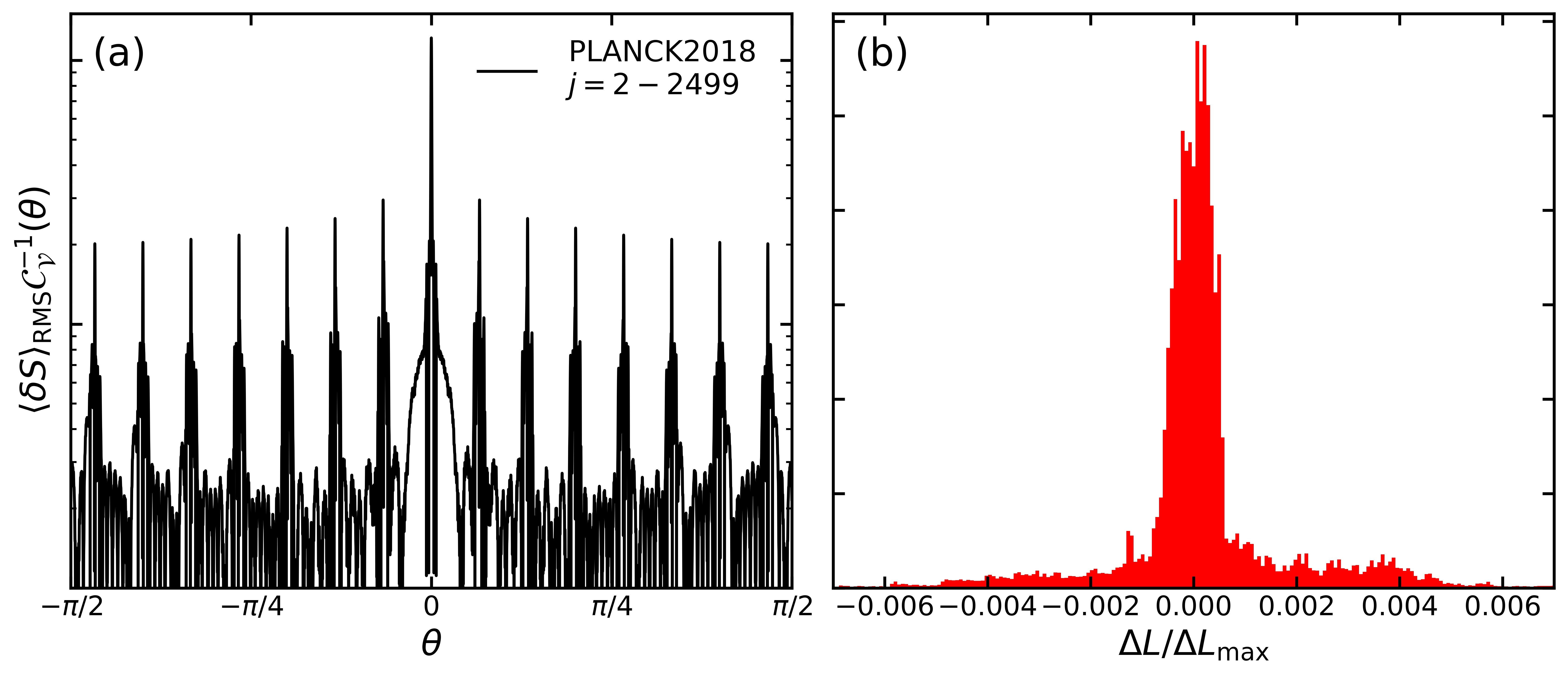}
	
	\caption{(a) Polar distribution of entanglement entropy of $N=10^5$ subsystems in the CMB sky according to PLANCK 2018 temperature fluctuation data (binned) \cite{aghanim2018planck}. The RMS entropy is proportional to entangling surface variation, $\frac{\langle \delta S \rangle_\mathrm{RMS}}{\mathcal{C}_\mathcal{V}}\propto \Delta L$. (b) Histogram distribution of the entangling surface from (a), normalized with respect to its maximum value. The corresponding average is at $\Delta L = 0.0011\Delta L_\mathrm{max}$.}
	\label{cmb_plot}
\end{figure}

Figure \ref{cmb_plot} depicts the polar distribution of  $\frac{\langle \delta S \rangle_\mathrm{RMS}}{\mathcal{C}_\mathcal{V}}\propto \Delta L$ for $N=10^5$ subsystems and their length histogram. In fig. \ref{cmb_plot}(a), the leading peak at $\theta=0$ corresponds to low $j=2-30$ and the dominant spikes correspond to the first peak of CMB at around $j=200$ (responsible to the flatness of the universe). The overall noisy profile comes from the acoustic oscillation of high $j>200$. The profile of the length distribution of $\Delta L$ in fig. \ref{cmb_plot}(b) deviates from Gaussian as the mean is located at about $0.0011\Delta L_\mathrm{max}$. Note that the distribution is very sharp; the variance is less than 0.1\% of $\Delta L_\mathrm{max}$. The relation between $\Delta L$ to visible and dark matter compositions needs further investigations.

Note that in applying our formalism to CMB data, we have interpreted the entanglement entropy as a thermal entropy via the entanglement temperature. The consequence of this implicit assumption is that the low-energy degrees of freedom in the Hamiltonian are responsible for the long-range entanglement in and near the ground state. This assumption is justified because there is a connection between entanglement entropy and thermal entropy in gapless systems whose low-energy physics is described by CFT \cite{swingle2013universal}. This also implies that the entanglement entropy of a local quantum system is at most of order $\log(L)$, although further clarification is needed. Nevertheless, from an observationalist (pragmatic) view, quantum and classical correlations are indistinguishable. Interactions with other degree of freedom would induce quantum-to-classical transition due to environmental decoherence, in a sense that the system can be entangled with its environment even without direct interaction. Non-linear effect in cosmology might induce large CMB non-Gaussianity in this context \cite{kiefer2009cosmological}. Combined with our findings, this may open a new perspective in the quantum aspects of cosmology.
\section{Conclusions and Discussions}
\label{sec:6}
In this paper we analyze a primordial highly-entangled quantum memory state, $|\psi\rangle ^{(j)}$, as a complicated superposition of a Fock space coming from the holographic jumps mechanism suggested by Gia Dvali. The entanglement entropy, i.e. von Newmann entropy is used to extract the entanglement measure of the quantum memory state. The holographic entanglement entropy or the Ryu-Takayanagi prescription is used to calculate the entanglement measure of a subsystem $A$ with respect to the rest, $A^c$. Since it is an N-body quantum system, we define the holographic multipartite entanglement entropy for the entanglement measure of $|\psi\rangle^{(j)}$. Some explicit calculations of the holographic multipartite entanglement entropy is presented in AdS$_3$/CFT$_2$ and AdS$_{d+1}$/CFT$_{d}$ systems.\\
\indent We found that the holographic multipartite entanglement entropy always reduces to the homogeneous mode ($j=0$) for the first-order entropy perturbation. This happen since we demand the total area of a compact CFT$_d$ system is identical for all $j$-modes. As a consequence, the distribution of the entropy perturbation, $\langle\delta S\rangle$, needs to be symmetric and hence a Gaussian distribution is a perfect match. We also found that, in the large N limit, the first-order entropy perturbation is proportional to the change of the subsystem size, i.e. $\delta S\propto\delta l$, and therefore the distribution of the entangling region size is also Gaussian. Using the entanglement thermodynamics formulation and $TdS$ equation, we also found that the change of entanglement entropy is proportional to the change of entanglement temperature with a proportionality constant, $\mathcal{C_{V}}$, called the entanglement heat capacity.\\
\indent Using these findings, we analyze the temperature distribution of the cosmic microwave background (CMB) sky. We construct the two-point correlation function of $\delta S$ and connect it to the temperature fluctuations of CMB via entanglement thermodynamics. We also choose $d=3$ which means that we are analyzing information that is spread out on a 2-dimensional spatial surface. In this prescription, we interpret the distribution of the entangling region size as the distribution of temperature fluctuations in the CMB sky. We obtain the distribution of the entanglement entropy, by matching them with the Planck 2018 CMB data, in Fig. \ref{cmb_plot}. This entropy method for analyzing CMB anisotropy also coincides with \cite{Minkov_2019, Helling_2006} that used the Wehrl entropy as the measure of deviation from classicality. We conclude that entanglement, via holographic entanglement entropy as the entanglement measure of highly-entangled primordial quantum memory states, might play a role in the analysis of CMB and this may open a new perspective in the quantum aspects of cosmology.\\
\indent There are several other interesting features of this work that might lead us to the future vast area of research. Since, in the holographic prescription, the change of entanglement entropy means geometric perturbation, the distribution of $\delta S$ might come from various arbitrary sources such as the existence of black holes or gravitational waves. The analysis of a deviation from Gaussian statistics of the $\delta S$ distribution also lights some insight into the role of entanglement in analyzing primordial non-Gaussianities. This insight might come from either changing the total CFT$_d$ size $\text{Area}(L)$ to be unique for some value(s) of $j$ or the possibility of $\rho_{A_1^{(j)}...A_N^{(j)}}$ being an initially mixed state. The latter shows that calculation with holographic entanglement of purification (and its multipartite counterpart) is needed.
\section*{Acknowledgements}
F. P. Z. would like to thank the Ministry of Research Republic of Indonesia and BRIN for financial support. H. L. P. and D. D. would like to thank the members of the Theoretical Physics Groups of Institut Teknologi Bandung for the hospitality.
\bibliographystyle{jhep}
\bibliography{bib}
\end{document}